\documentclass[pra,twocolumn,showpacs,superscriptaddress,byrevtex]{revtex4}

\usepackage{graphicx}
\usepackage{epsfig}
\usepackage{dcolumn}
\usepackage{bm}
\usepackage{amsmath}
\usepackage{amssymb}
\usepackage[dvips]{hyperref}
\hypersetup{colorlinks=true,linkcolor=blue,citecolor=blue,urlcolor=black}

\newcommand{\ket}[1]{\ensuremath{|\,{#1}\,\rangle}}

\renewcommand{\vec}[1]{\mbox{\boldmath{\ensuremath{{#1}}}}}

\newcommand{\itgf}[1]{\ensuremath{\int\!\!d{#1}\,}}

\newcommand{\sinc}{\ensuremath{\mbox{\hspace{1.3pt}sinc}\,}}

\begin{document}


\title{Fast entanglement detection for unknown states of two spatial qutrits}

\author{G. Lima}
\email{glima@udec.cl}
\affiliation{Center for Optics and Photonics, Universidad de Concepci\'{o}n, Casilla 4016, Concepci\'{o}n, Chile}
\affiliation{Departamento de F\'{i}sica, Universidad de Concepci\'{o}n, Casilla 160-C, Concepci\'{o}n, Chile}
\author{E. S. G\'{o}mez}
\affiliation{Center for Optics and Photonics, Universidad de Concepci\'{o}n, Casilla 4016, Concepci\'{o}n, Chile}
\affiliation{Departamento de F\'{i}sica, Universidad de Concepci\'{o}n, Casilla 160-C, Concepci\'{o}n, Chile}
\author{A. Vargas}
\affiliation{Center for Optics and Photonics, Universidad de Concepci\'{o}n, Casilla 4016, Concepci\'{o}n, Chile}
\affiliation{Departamento de Ciencias F\'{i}sicas, Universidad de La Frontera, Temuco, Casilla 54-D, Chile}
\author{R. O. Vianna}
\affiliation{Departamento de F\'{i}sica, ICEx, Universidade Federal de Minas Gerais, Belo Horizonte, MG, Brazil}
\author{C. Saavedra}
\affiliation{Center for Optics and Photonics, Universidad de Concepci\'{o}n, Casilla 4016, Concepci\'{o}n, Chile}
\affiliation{Departamento de F\'{i}sica, Universidad de Concepci\'{o}n, Casilla 160-C, Concepci\'{o}n, Chile}
\date{\today}



\begin{abstract}
We investigate the practicality of the method proposed by Maciel {\em et al.} [Phys. Rev. A. {\bf 80}, 032325 (2009)] for detecting the entanglement of two spatial qutrits ($3$-dimensional quantum systems), which are encoded in the discrete transverse momentum of single photons transmitted through a multi-slit aperture. The method is based on the acquisition of partial information of the quantum state through projective measurements, and a data processing analysis done with semi-definite programs. This analysis relies on generating gradually an optimal entanglement witness operator, and numerical investigations have shown that it allows for the entanglement detection of \emph{unknown} states with a cost much lower than full state tomography.
\end{abstract}

\maketitle


\section{Introduction}

One of the main goals of quantum information theory is to find general and simple procedures that determine whether a composite system is entangled or not \cite{Horodeckis}. The simplest example is the entanglement detection of pure bipartite states of qubits ($2$-dimensional quantum states). In this case, the entanglement of the compound state can be determined directly from the reduced density matrixes \cite{Bennett,Wootters}. This approach can be generalized for distinct types of quantum states \cite{Bennett2,Wootters2,Rungta}, and they all rely on state dependent nonlinear functions that characterize entanglement quantitatively. These functions can be evaluated once one has the knowledge of the compound density operator.

Unfortunately, the experimental determination of composite states can require an extremely fastidious work \cite{Haffner,Weinfurter}.  An unknown $D$-dimensional quantum state (qudit) is represented by a $D \times D$ density operator, which requires $D^2 - 1$ independent real parameters for its specification. The state reconstruction is done with the technique of quantum tomography (QT), where these parameters are associated, usually, with an overcomplete set of observables that are evaluated on single copies of the quantum system in consideration \cite{Fano,James1}. Besides of the large amount of measurements necessary in the QT, it may also require sophisticated programs for doing the unavoidable numerical optimization of the acquired data \cite{James2}. Depending on the dimension of the composite system, this optimization may be even impossible with today's technology \cite{James2}.

Therefore, it is legitimate to ask if there are more practical schemes that allow for the entanglement detection of composite states, which do not require the full state reconstruction. One possibility is to consider the so-called collective measurements \cite{Acin,Bovino,Mintert1,Steve,Mintert2,Weinfurter2,Aolita}, where the idea is to detect the entanglement by measuring observables that act on simultaneously available copies of the composite state. However, the simultaneous generation of copies of a quantum state may not be possible in some circumstances, and another solution for the entanglement detection is the use of entanglement witnesses operators \cite{HorodeckiEW,Terhal}. These operators are observables ($W$) with non-negative mean values for all separable states, such that the measurement of a negative mean value [$Tr(W\rho_{AB}) < 0$] witness the entanglement of the compound state $(\rho_{AB})$. It is also worth to note that, recently, a link between collective measurements and single-copy-entanglement-witnesses has been presented \cite{Mintert3}.

In general, there is no universal witness operator (see, however, Ref \cite{HorodeckiUni} for a particular case) and one must assume a prior knowledge of the composite state for constructing the optimal EW operator \cite{Mataloni,Bourennane}. Nevertheless this approach is of limited application, since this prior knowledge of the state may not be available. Moreover, the witness operator will give reliable information only if the considered state is detected. Otherwise one can not distinguish whether the state was separable or contained a distinct type of entanglement.

The efficient construction of the entanglement witnesses is also another important drawback for this technique of entanglement detection. There are many distinct techniques presented \cite{Acin2,Reinaldo,Toth1,Guhne2,Toth2}, but the optimal way for witnessing the entanglement of a given state is still an unsolved problem.

For circumventing the practical problem of witnessing the entanglement of \emph{unknown} states, there has been, recently, the development of a numerical and interactive method, where the witness operator is constructed and evaluated \emph{while} the experimental data is collected \cite{Reinaldo2,Reinaldo3}. For detecting the entanglement of a compound state from an incomplete
set of measurements, one can also use the techniques of \cite{Horodeckis2,Guhne3}. The need for such procedures arises often in real applications, and in this work we present the first experimental demonstration of the method described in \cite{Reinaldo2,Reinaldo3}. It is interesting to note that numerical investigations have shown that it allows for the entanglement detection with a cost much lower than QT.

For testing the method of Maciel {\em et al.} \cite{Reinaldo2,Reinaldo3}, we generated bipartite entangled states of $3$-dimensional quantum systems (qutrits). The qutrits are encoded in the discrete transverse momentum of single photons transmitted through a multi-slit aperture, and we refer to them as \emph{spatial qutrits} \cite{Leo}. Pure entangled states with distinct degrees of entanglement \cite{Bennett,Wootters} were used to study the method when it is applied to a system with a variable resource, namely entanglement. We also considered different measurement settings for each state generated, and this allows one to verify the experimental convergence of the method under distinct scenarios.

The experimental test of a theoretical method is important to determine the practicality and limitations of the method in actual implementations. Besides, it is important to note that the work presented here can also be interpreted as a new technique for estimating the entanglement of transverse spatially correlated quantum systems, whose characterization has been the theme of recent investigations \cite{Leo2,Glima0,Japas,Exter,Monken}. As it was shown in \cite{Guhne3,Eisert}, it is possible to associate lower bounds for some entanglement measures, given that the mean value of a witness operator has been determined.

\section{Experiment}

\subsection{Generating the spatial qutrits}

\begin{figure}[t]
\centerline{\includegraphics[width=0.47\textwidth]{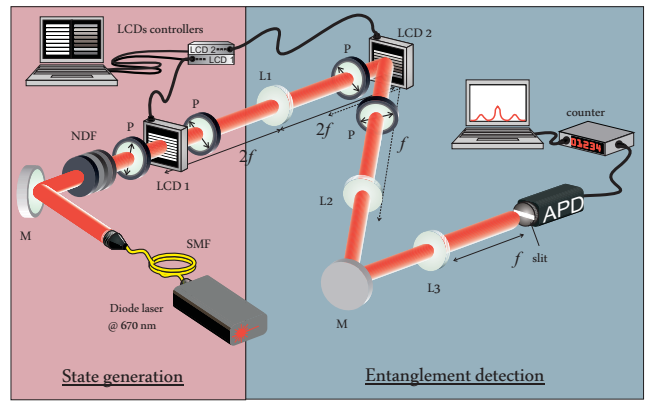}}
\caption{Experimental Setup. See the main text for
details.} \label{Fig:Setup}
\end{figure}

In Fig.~\ref{Fig:Setup}, it is represented the setup used for the generation and
entanglement detection of the spatial qutrits states.
A pigtailed single mode diode laser, operating at $670$~nm, is attenuated to the single photon level with absorptive neutral density filters (NDF). The particle behavior of the light was confirmed by using an avalanche
photodetector (APD) to register the transmitted photons. After $2$~m
of free propagation (which is the overall extension of our
experimental setup), the maximal single count rate recorded was
$50000$~$s^{-1}$. Taking into account the coherence length of the
diode laser and the APD properties, this shows that, on average,
fewer than $10^{-3}$ photons is presented in the experiment at
any given moment.

After this initial test, the transmitted photons
are sent to a transmissive spatial light modulator working for amplitude-only modulation. It is
composed of two polarizers (P) and a twisted nematic liquid crystal
display (LCD). The polarizers are placed close to the LCD panel ($0.5$~cm away from each side). The LCD is connected to a PCB interface which allows its control from a computer. This SLM is addressed with a
$9$-multi slit. The slit's width is $2a=104$~$\mu$m and the distance between two consecutive slits is $d=208$~$\mu$m. In this case, the overall SLM transmission is $20\%$. The state of the single photons
transmitted through this SLM can be written as~\cite{Leo,Glima}
\begin{equation} \label{State}
\ket{\Psi} = \sum_{l=-4}^4 \alpha_l\ket{l},
\end{equation} where the state $\ket{l}$ is defined by $\ket{l} \equiv \sqrt{\frac{a}{\pi}} \itgf{q} e^{-
iqld}\sinc(qa)\ket{1q}.$ It represents the state of the photon transmitted by the $l$th-slit of the SLM.
$\ket{1q}$ is the Fock state of a photon with the transverse wave
vector $\vec{q}$. These states $\ket{l}$ form an orthonormal basis
in the $9$-dimensional Hilbert space of the transmitted photons and,
therefore, they are used to define the logical spatial states. The state of Eq.~(\ref{State}) represents a $9$-dimensional quantum state that is encoded in the discrete transverse propagation modes of the transmitted photons. Alternatively, it can also be thought as a composite system of two qutrits, which can be entangled or not. For seeing this, one just need to consider a new label for the $\ket{l}$ states, like for example $\ket{l}\rightarrow \ket{ij}$,
such that: $\ket{-4}\rightarrow \ket{00}$; $\ket{-3}\rightarrow \ket{01}$;
$\ket{-2}\rightarrow \ket{02}$; $\ket{-1}\rightarrow \ket{10}$; $\ket{0}\rightarrow \ket{11}$; $\ket{1}\rightarrow \ket{12}$; $\ket{2}\rightarrow \ket{20}$; $\ket{3}\rightarrow \ket{21}$ and $\ket{4}\rightarrow \ket{22}$. In this case, the state of the transmitted photons is given by
\begin{equation} \label{Stateij}
\ket{\Psi} = \sum_{i,j=0}^2 \alpha_{ij}\ket{ij}.
\end{equation} These spatial qutrits are encoded in the transverse spatial modes of a single particle, and therefore they can not be used to test quantum non-locality \cite{Bell,CHSH}. Nevertheless, they can be used to study other fundamental aspects of quantum mechanics \cite{Spreeuw}.

To generate spatial qutrits states with different degrees of entanglement, it is necessary to modify the coefficients $\alpha_{ij}$ of Eq.~(\ref{Stateij}). These coefficients are dependent on the spatial
profile and the wavefront curvature of the laser beam at the SLM plane. They are also dependent on the slit's transmissions addressed in the SLM, and as it was shown in \cite{Glima}, the spatial light modulator
can be used to modify, independently, each slit transmission. This modulation is done by changing the grey level of the LCD at the points where the slits are addressed, and thus, it allows for a controlled generation of
distinct types of spatial qutrits entangled states. This manipulation has been demonstrated to be coherent, such that the modified states
can still be described by a coherent superposition~\cite{Glima}.

\begin{figure}[t]
\centerline{\includegraphics[width=0.52\textwidth]{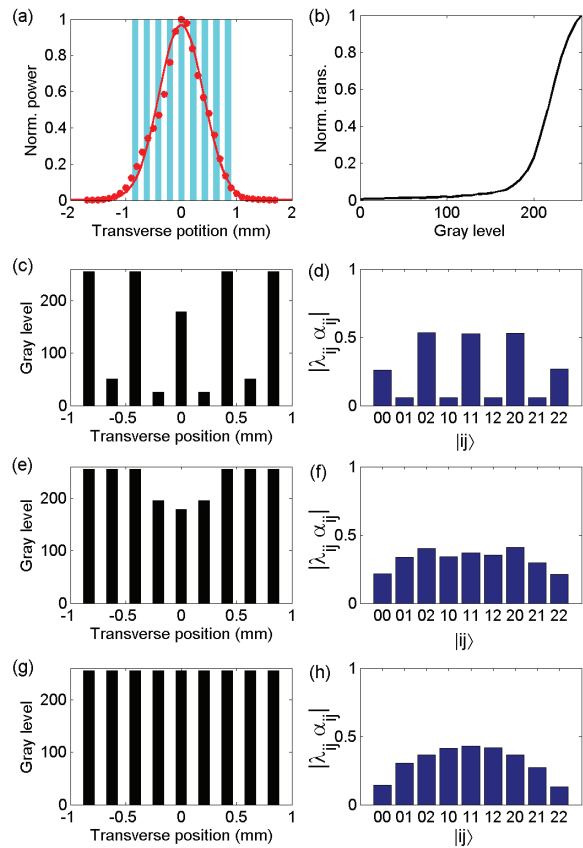}}
\vspace{-0.5cm}
\caption{The expected spatial qutrits states. In (a) it is
shown a comparison between the laser beam spatial profile and the
$9$-multi slit aperture addressed in the first SLM. In (b) it is shown the dependence of
the SLM transmission and the grey level of its LCD. In Fig.~\ref{Fig:StateExpc}(c), Fig.~\ref{Fig:StateExpc}(e) and Fig.~\ref{Fig:StateExpc}(g) it is shown the modulations used for generating three distinct types of spatial qutrits states. The corresponding expected amplitudes of these states are shown in Fig.~\ref{Fig:StateExpc}(d), Fig.~\ref{Fig:StateExpc}(f) and Fig.~\ref{Fig:StateExpc}(h), respectively.} \label{Fig:StateExpc}
\end{figure}

In our work, the qutrits states were generated by sending a well collimated beam to the first spatial light modulator. In this case, the coefficients $\alpha_{ij}$ are expected to be real, and the initial state of Eq.~(\ref{Stateij}) can be modified by the SLM to

\begin{equation}
\ket{\Psi}_{mod1} = \frac{1}{\sqrt{N}}
\sum_{i,j=0}^{2} \lambda_{ij} \alpha_{ij} \ket{ij},
\label{StateMOD1}
\end{equation} where \protect{$\lambda_{ij}\equiv\sqrt{t_{ij}}$}, with $t_{ij}$ representing the transmission of the $ij$-slit. $N$ is a normalization constant.
In Fig.~\ref{Fig:StateExpc}(a) it is shown a comparison between the laser spatial profile at the plane of the SLM and the 9-multi slit being
addressed on it. This graph shows how each slit of the array was initially illuminated. In Fig.~\ref{Fig:StateExpc}(b) one can see the dependence of the SLM transmission and the grey level of its LCD. In Fig.~\ref{Fig:StateExpc}(c), Fig.~\ref{Fig:StateExpc}(e) and Fig.~\ref{Fig:StateExpc}(g) it is shown the modulations used for generating three distinct types of spatial qutrits states. The corresponding expected amplitudes of these states are shown in Fig.~\ref{Fig:StateExpc}(d), Fig.~\ref{Fig:StateExpc}(f) and Fig.~\ref{Fig:StateExpc}(h), respectively. These amplitudes are calculated by taking into account the spatial laser profile and the transmissions of the slits being addressed in the SLM. As it will be discussed on next section, these states have different degrees of entanglement. The first state generated is a highly entangled state of two spatial qutrits, while the second and the third states are generated by reducing gradually the degree of entanglement between the qutrits.

\subsection{Characterizing the states generated}

\begin{figure*}[t]
\centerline{\includegraphics[width=0.9\textwidth]{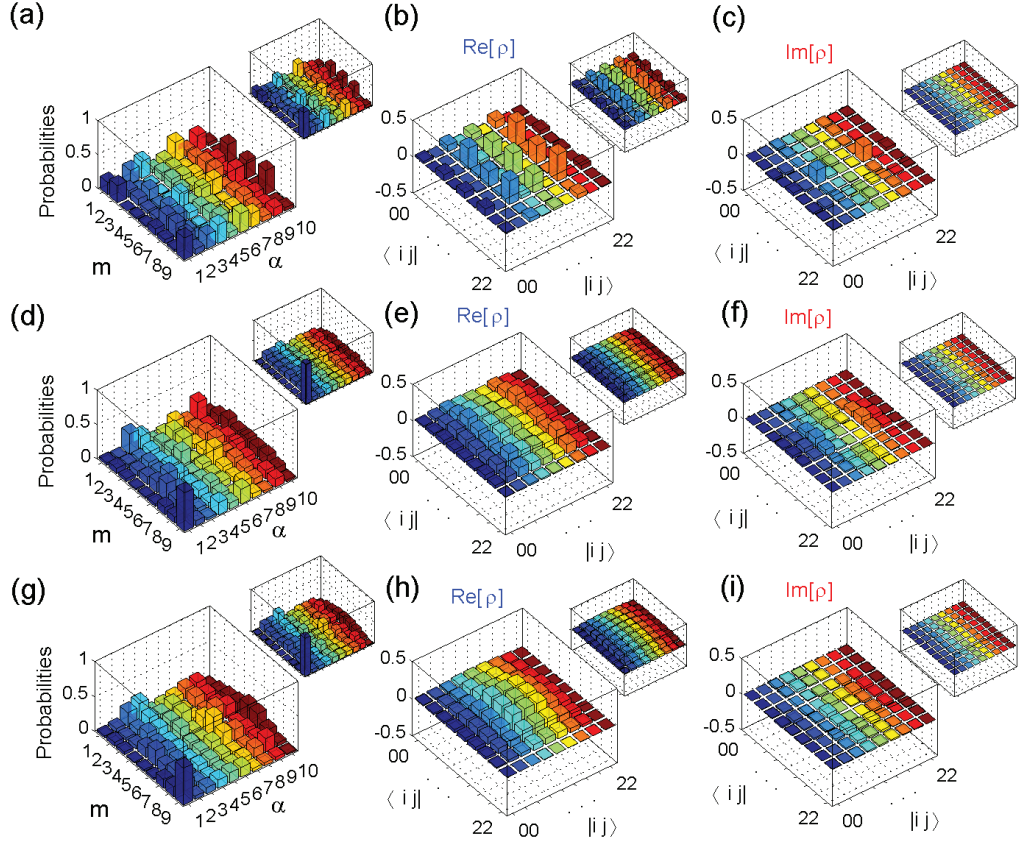}}
\vspace{-0.5cm} \caption{Reconstructed states. In (a), (d) and (g) it is shown the recorded probabilities and the expected ones (insets) that are calculated from the states given in Fig.~\ref{Fig:StateExpc}(d), Fig.~\ref{Fig:StateExpc}(f) and Fig.~\ref{Fig:StateExpc}(h), respectively. In (b) [(c)], (e) [(f)] and (h) [(i)] it is shown the real (imaginary) parts of the corresponding reconstructed density operators in the logical base.  On the insets it is shown the expected parts of these density operators.} \label{Fig:QutritsRecons}
\end{figure*}

Even though we can estimate which are the spatial qutrits entangled states generated in our experiment (See Fig.~\ref{Fig:StateExpc}), this is not enough for demonstrating that the method of \cite{Reinaldo2,Reinaldo3} allows for a fast detection of their entanglement. For doing such study it is first necessary to completely determine these states, so that we know precisely what we should obtain in the test.

In this work we use the recently proposed technique of Ref~\cite{Glima2} for performing the quantum tomography of the spatial qutrits states. It is based on projecting these states onto the vectors of the mutually unbiased bases (MUBs), and it has the advantage that the overcomplete set of vectors used for the reconstruction has a number of projections that is minimal \cite{Ivanovic,Wootters3}. In the MUBs, any two vectors of different bases have the same overlap's absolute value. There exist $10$ MUBs for a $9$-dimensional Hilbert space. The existence of these MUBs and
constructive routines for obtaining them, has been given by Wootters in \cite{Wootters3}. The density operators of the spatial qutrits states can be represented in terms of the MUBs as \cite{Ivanovic}
\begin{equation}
\rho=\sum_{\alpha=1}^{10}\sum_{m=1}^9
p_m^{(\alpha)}\Pi_m^{(\alpha)}-I, \label{OpeMUB}
\end{equation} where $\Pi_m^{(\alpha)}\equiv |\psi_m^{(\alpha)}\rangle
\langle\psi_m^{(\alpha)}|$, and
$p_m^{(\alpha)}=Tr(\rho\Pi_m^{(\alpha)})$ is the probability for
projecting $\rho$ onto the $m$-th state $|\psi_m^{(\alpha)}\rangle$
of the $\alpha$ MUB. We note that the states $|\psi_m^{(\alpha)}\rangle$ are given explicitly in Ref~\cite{Wootters3}. $I$ represents the $9$-dimensional unity operator. Thus, the QT based on the MUBs of the spatial qutrits is implemented by measuring the probabilities $p_m^{(\alpha)}$ of the $90$ corresponding projectors ($\Pi_m^{(\alpha)}$).

For projecting the generated spatial qutrits states onto the vectors of the MUBs, it is necessary to consider a second SLM and a spatial filtering process as it is described in details in \cite{Glima2}. Briefly, a second SLM is set with its liquid crystal display (LCD2) placed at the plane of image formation of the first LCD. In our case we used a $7.5$~cm focal length lens (L1) (See Fig.~\ref{Fig:Setup}). This SLM must be configured for doing phase-only modulation and then it is used to modulate the phase of the transmitted light, independently, at each point of slit image formation. This is done by changing the gray level of the LCD$2$, independently, at these points. At the
plane of image formation, the diffracted photon is again described by Eq~(\ref{StateMOD1}) \cite{GLima3} and, therefore, the modulation being done by the second SLM allows for the modification of the imaginary parts of the spatial qutrits states. After the reflection on this SLM, the light is collimated and focused by a set of lenses (L2 and L3) at the detection plane, where a point-like detector (APD) is used to record the
single photons at the \emph{center} of the interference patterns formed. This detector is composed of
a conventional avalanche photo-counting module with a slit of $20$~$\mu$m in front of it. The spatial filtering being introduced by the set of lenses and the point-like detector, together with the phase modulations of the second SLM, allows for projecting the spatial qutrits states onto the $|\psi_m^{(\alpha)}\rangle$ MUBs vectors of their $9$-dimensional Hilbert space \cite{Glima2}.

The recorded probabilities $p_m^{(\alpha)}$, for the first, second and third state generated in our experiment are shown in Fig.~\ref{Fig:QutritsRecons}(a), Fig.~\ref{Fig:QutritsRecons}(d) and Fig.~\ref{Fig:QutritsRecons}(g), respectively. On the insets of these figures it is shown the expected probabilities, which are calculated from the states given in Fig.~\ref{Fig:StateExpc}(d), Fig.~\ref{Fig:StateExpc}(f) and Fig.~\ref{Fig:StateExpc}(h), respectively. In Fig.~\ref{Fig:QutritsRecons}(b) and Fig.~\ref{Fig:QutritsRecons}(c) it is shown the real and the imaginary parts of the reconstructed density operator of the first state. On the insets it is shown the expected real and imaginary parts. For obtaining this density operator we replace the values of the probabilities $p_m^{(\alpha)}$ shown in Fig.~\ref{Fig:QutritsRecons}(a) into Eq.~\ref{OpeMUB}, and then optimize the resulting matrix to guarantee that it is positive semi-definite \cite{James1}. For doing this optimization we considered the ``forced purity'' approach, which has been shown to be sufficient when one is dealing with high-dimensional quantum systems of low-entropy \cite{James2}. As it was shown in \cite{Glima2}, this is exactly our case. The ``forced purity'' numerical technique generates a pure quantum state for which the mean values of the measurements considered are very close to the ones obtained in the experiment. The fidelity \cite{Jozsa} of the first reconstructed state with the expected one [Fig.~\ref{Fig:StateExpc}(d)] was $F_{state1}=0.84\pm0.04$. The real and the imaginary parts of the density operator obtained for the second (third) state are shown in Fig.~\ref{Fig:QutritsRecons}(e) [Fig.~\ref{Fig:QutritsRecons}(h)] and Fig.~\ref{Fig:QutritsRecons}(f) [Fig.~\ref{Fig:QutritsRecons}(i)], respectively. The expected parts of these density operators are also shown on the insets of these figures. For the second state, the fidelity between the obtained density operator and the expected one was $F_{state2}=0.87\pm0.04$. For the third state we got a fidelity of $F_{state3}=0.92\pm0.04$. The high value of the fidelities between the reconstructed states and the expected ones is usually accepted as a demonstration that the experimental setup is working as it is intended to, so that we may use it for further studies.

Next we investigate the numerical technique of \cite{Reinaldo2,Reinaldo3}. This technique is dependent of the degree of entanglement of the composite system and so, it is important that we determine also the degree of entanglement of the reconstructed states of our experiment. For calculating the degree of entanglement of the spatial qutrits states we use the \emph{I concurrence} introduced by Rungta \emph{et al.} \cite{Rungta}, which is a generalization of the concurrence \cite{Wootters} to arbitrary dimensional bipartite composite systems. For pure bipartite systems, it is dependent only on the purity of the marginal density operators
\begin{equation} \label{Concg}
(\mathcal{C}_{AB}^{(n)})^2=2\left[1-tr(\rho_k^2)\right],
\end{equation} where $k=A,B$ denotes the subsystems. In this case of pure bipartite systems, the \emph{I concurrence} covers the range of \cite{Bergou}
\begin{equation}
0\leq(\mathcal{C}_{AB}^{(n)})^2\leq\frac{2(n-1)}{n},
\end{equation} where $n\equiv min[n_A,n_B]$, with $n_A$ and $n_B$ defining the subsystems dimensions. In the case of pure spatial qutrits states we have that $0\leq(\mathcal{C}_{AB}^{(3)})^2\leq\frac{4}{3}$. From Eq.~(\ref{Concg}) we can calculate the \emph{I concurrence} of the reconstructed spatial qutrits states given in Fig.~\ref{Fig:QutritsRecons}. The calculated values are shown in Tab.~\ref{tab1}.

\begin{table}[h]
\begin{center}
\begin{tabular}{c|c} \hline \hline
\multicolumn{2}{c}{\emph{I concurrence}} \\
\hline \hline
State &$\left(\mathcal{C}_{AB}^{(3)}\right)^2$  \\
\hline
1 &$1.20$ \\
\hline
2 &$0.45$ \\
\hline
3 &$0.27$
\end{tabular}
\caption{\emph{I concurrence} of the reconstructed spatial qutrits states.} \label{tab1}
\end{center}
\end{table}

\subsection{Fast entanglement detection for unknown spatial qutrits states}

\subparagraph{Overview of the method --} The method of Maciel \emph{et al.} \cite{Reinaldo2,Reinaldo3}, can be seen as a variational approach that allows one to estimate some physical properties of an unknown quantum system. It requires the acquisition of partial information of the state of the system, obtained through projective measurements, and it consists of a interactive routine, where semi-definite programs \cite{semiprograms} are used to gradually estimate the expectation value of an observable while the data is collected.

This estimation has been showed to be optimal and reliable \cite{Reinaldo3}. Optimal in the sense that it converges to the correct mean value of the observable in consideration. Reliable, because the program will not attribute to a system an upper value for its physical property. The obtained expectation value will always be smaller in absolute value than the true one. This is an important property, specially, when one is detecting the entanglement of the system with entanglement witnesses. In this case, the entanglement of the system will never be overestimated.

The basic idea of the method is the following. Suppose some projectors have been measured on the unknown quantum state. These projectors can be, for example, some of the projectors we used above ($\Pi_m^{(\alpha)}$), in the case one is dealing with spatial qutrits states. Once these projectors have been measured, one will have determined the corresponding projection probabilities (some of the probabilities $p_m^{(\alpha)}$ in our case), and it is then possible to write the expectation value of an observable as a function of two terms: one which is a function of the known probabilities, and one which depends on the unknown ones \cite{Reinaldo3}. Thus, the idea is to perform a variational estimation of the expectation value, by doing an appropriate minimization of the term which depends on the unknown probabilities. This problem consist of a linear optimization problem, which can be solved efficiently with semi-definite programs.

Since the method allows for the estimation of observables, it can be used for doing the entanglement detection of composite states with the technique of entanglement witnesses operators \cite{Reinaldo2}. To build the witness operator, the strategy used is the one of Ref \cite{Reinaldo}, which allows for the construction of an optimal trace one witness. Besides, the method can be used for the entanglement detection of unknown states. This is possible because the witness operator is built according to a ``guess'' state, $\tilde{\rho}$, which is efficiently inferred from the projective measurements already performed in the composite system \cite{Reinaldo3}. When the number of projective measurements increases, we have that $\tilde{\rho} \rightarrow \rho$ (where $\rho$ represents the real/actual state of the system). When the measurements are equivalent to the full QT, the witness operator corresponds to the optimal one for the compound system.

\begin{figure}[t]
\vspace{0.5cm}
\centerline{\rotatebox{-90}{\includegraphics[width=0.39\textwidth]{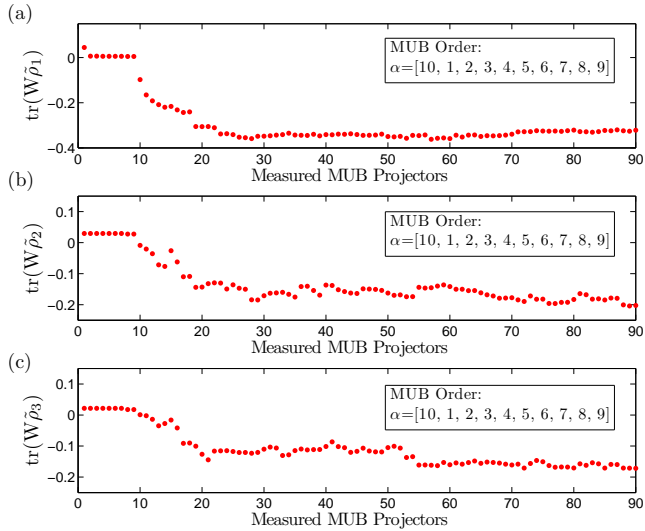}}}
\caption{Entanglement detection for the generated spatial qutrits states, without assuming any knowledge of them. In (a) it is shown the results obtained for the first state generated in our experiment. In (b) [(c)] it is shown the results obtained for the second (third) state generated. See the main text for details.} \label{Fig:FastEnt}
\end{figure}

\subparagraph{Fast entanglement detection for unknown spatial qutrits states --} Now we show how the program was used for doing the fast entanglement detection of the states generated in our experiment. No prior knowledge of these states is assumed. The data used in the analysis is the same acquired during the state reconstruction. The recorded probabilities for these states are shown in  Fig.~\ref{Fig:QutritsRecons}(a), Fig.~\ref{Fig:QutritsRecons}(d) and Fig.~\ref{Fig:QutritsRecons}(g). While the data was being collected, the program was used to detect the entanglement of the generated states. In Fig.~\ref{Fig:FastEnt}(a), Fig.~\ref{Fig:FastEnt}(b) and Fig.~\ref{Fig:FastEnt}(c) it is shown the obtained results for the first [$\left(\mathcal{C}_{AB}^{(3)}\right)^2=1.20$], second [$\left(\mathcal{C}_{AB}^{(3)}\right)^2=0.45$] and the third state [$\left(\mathcal{C}_{AB}^{(3)}\right)^2=0.27$], respectively. The order of the measurements performed is indexed with $\alpha=[10,1,2,3,4,5,6,7,8,9]$. This means that the first 9 projective measurements implemented correspond to the projectors of the $10$th-MUB, the 10th to 18th measurements correspond to the projections onto the $1$st-MUB and etc... The $10$th-MUB, as one can see in Fig.~\ref{Fig:QutritsRecons}, corresponds to the logical base of the qutrits Hilbert space, and this is why it was measured first. This was done according to the procedure described in \cite{Glima2}.

From Fig.~\ref{Fig:FastEnt} one can observe that the entanglement was always detected much faster than QT. While QT requires a total of 90 measurements to determine the entanglement of the generated states, the method of \cite{Reinaldo2,Reinaldo3} allowed the entanglement detection already on the second measured MUB. This happened independently of the degree of entanglement between the spatial qutrits.

Another important observation to be done is regarding the convergence observed for each state. For the first state (Fig.~\ref{Fig:FastEnt}(a)), the convergence was faster. 23 measurements were necessary for a reasonable estimation of the witness expectation value. For the second state (Fig.~\ref{Fig:FastEnt}(b)), which is less entangled, the system required around 30 measurements. For the last state (Fig.~\ref{Fig:FastEnt}(b)), which has the lowest $I$ concurrence, more than 55 measurements were necessary. This is important because as it was shown in \cite{Guhne3,Eisert}, it is possible to associate lower bounds for some entanglement measures, given that the mean value of a witness operator has been determined. Thus, we may say that the method of \cite{Reinaldo2,Reinaldo3} also allowed for the fast estimation of the entanglement of the spatial qutrits states. Moreover, the behavior observed is exactly what one should expected, since it should be, in principle, easier to estimate the entanglement of states that are more entangled.

\begin{figure}[b]
\vspace{-1cm}
\centerline{\includegraphics[height=10cm,width=10cm]{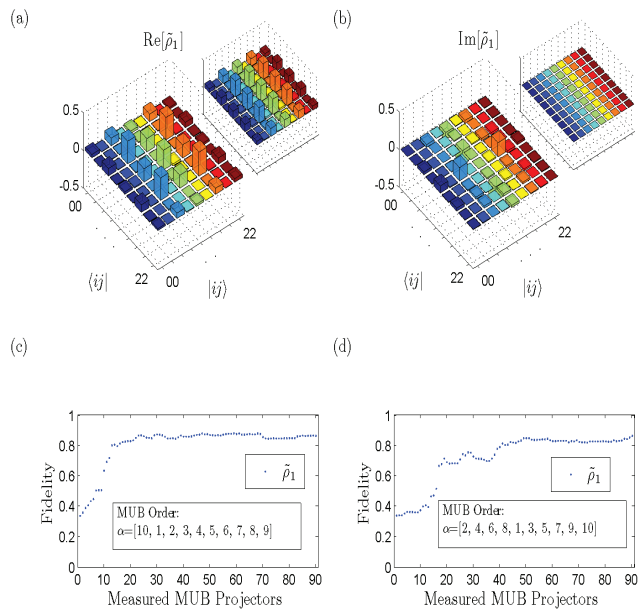}}
\vspace{-0.5cm}
\caption{State reconstruction for the first entangled state of spatial qutrits. In (a) and (b) it is shown a comparison between the final obtained state $\tilde{\rho}_1$, and the expected one (insets) calculated from the amplitudes shown in Fig.~\ref{Fig:StateExpc}(d). In (c) it is shown the fidelities between these states while the measurements were being performed. In (d) it is shown the fidelities between these states when the measurements are taken in a distinct order.} \label{Fig:RecRein}
\end{figure}

\subparagraph{Further analysis --} As it is discussed in details in \cite{Reinaldo3}, the numerical technique we are investigating turns out to be also an algorithm that allows for QT with an incomplete set of observables. As we mentioned above, this is done with a numerical estimation of the state of the system, that takes place after some projective measurements have been performed. In Fig.~\ref{Fig:RecRein}, it is shown for the first state of our experiment, a comparison between the state generated by the method of Maciel \emph{et al.}, and its expected form calculated from the amplitudes given in Fig.~\ref{Fig:StateExpc}(d). One can see that the reconstruction presented a high value of fidelity, and this can be seen as an evidence of the good quality of the reconstruction done. \emph{The same behavior has been observed for the other states}. The fidelities between the final density operators obtained with this method, for the second and third state generated in the experiment, while compared with their expected form, were $78\%$ and $82\%$, respectively.

It is interesting to note in Fig.~\ref{Fig:RecRein}(c) and Fig.~\ref{Fig:RecRein}(d) that the convergence of the method depends on the measurement settings. In Fig.~\ref{Fig:RecRein}(c) it is shown the fidelities obtained while the measurements were being performed. Figure~\ref{Fig:RecRein}(d) was generated by applying the method to the data after it has been organized in a different order. One can see that the convergence was faster in Fig.~\ref{Fig:RecRein}(c).

\begin{figure}[h]
\vspace{0.5cm}
\centerline{\rotatebox{-90}{\includegraphics[height=8.5cm,width=9cm]{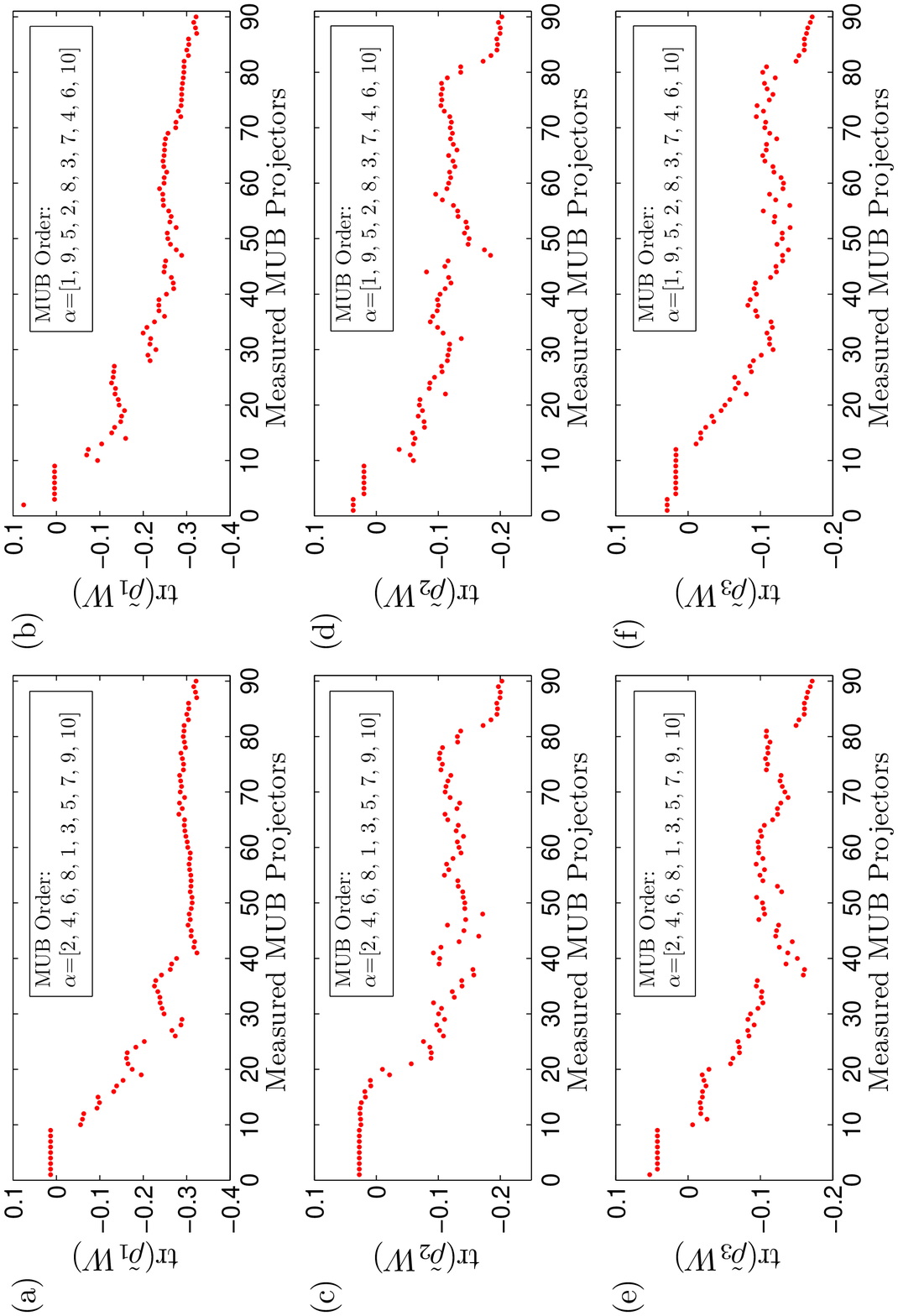}}}
\caption{Entanglement detection for the generated spatial qutrits states, without assuming any knowledge of them.  In (a) and(b) we have the entanglement detection for the first state, when the measurements are considered in a different order at the program. In (c) and (d) we have the entanglement detection for the second state. In (e) and (f) we have the same type of analysis for the third state. See the main text for details.} \label{Fig:FastEnt2}
\end{figure}

Since we have seen that the convergence of $\tilde{\rho}$ changes with the order of the measurements performed, it is interesting to have a look on how the expectation values of the witnesses operators are affected. This is shown in Fig.~\ref{Fig:FastEnt2}. In Fig.~\ref{Fig:FastEnt2}(a) and Fig.~\ref{Fig:FastEnt2}(b) we have the entanglement detection for the first state, when the measurements are considered in a different order at the program. In Fig.~\ref{Fig:FastEnt2}(c) and Fig.~\ref{Fig:FastEnt2}(d) we have the entanglement detection for the second state. In Fig.~\ref{Fig:FastEnt2}(e) and Fig.~\ref{Fig:FastEnt2}(f) we have the same type of analysis for the third state. One can see that even though the convergence of the method is lower than before (Fig.~\ref{Fig:FastEnt}), it can still detect the entanglement of the generated states much faster than QT.

\section{Conclusions}

In this work we have presented a detailed investigation of the technique recently introduced by Maciel \emph{et al.} \cite{Reinaldo2,Reinaldo3}, which allows for the fast entanglement detection of unknown states of composite systems. For testing the method we considered entangled states of two qutrits, which were generated with distinct degrees of entanglement. This allowed the experimental test of the method under distinct scenarios. The qutrits states were encoded on the discrete transverse modes of single photons transmitted through a diffractive aperture. Our results show that the method of \cite{Reinaldo2,Reinaldo3} can indeed provide fast entanglement detection even when experimental errors, such as misalignment and imperfections, are present. Even though we have studied the behavior of the method using only pure entangled states, it has been shown to be also valid for mixed states. It is also important to note that the method can be applied to any set of projective measurements used for doing the quantum tomography in previous experiments. The work presented here can also be seeing as a new technique for doing the entanglement estimation of spatially correlated quantum systems.

\begin{acknowledgments}
G. L. acknowledges L. Neves for fruitful discussions and R. Guzm\'{a}n for helping with the initial calibration of the SLMs. This work was supported by Grants Milenio ICM P06-067F and FONDECYT~11085055. A. V. acknowledges grant DI09-0045 of Universidad de La Frontera.
\end{acknowledgments}

\end{document}